\documentclass[11pt]{article}
\usepackage{graphicx}
\usepackage{amsmath}
\usepackage{amssymb}
\usepackage{caption2}
\begin{document}
\bibliographystyle{prsty}
\begin{center}
{\large {\bf \sc{  Analysis of  the $\Omega_b(6316)$,  $\Omega_b(6330)$,  $\Omega_b(6340)$ and $\Omega_b(6350)$  with QCD sum rules }}} \\[2mm]
Zhi-Gang Wang \footnote{E-mail:zgwang@aliyun.com.  }     \\
 Department of Physics, North China Electric Power University,
Baoding 071003, P. R. China
\end{center}

\begin{abstract}
In this article, we introduce an explicit P-wave  to construct three-quark currents to study the P-wave $\Omega_b$ states with the full  QCD sum rules. The predicted masses have a hierarchy if the same parameters are chosen and favor assigning the $\Omega_b(6316)$, $\Omega_b(6330)$, $\Omega_b(6340)$ and $\Omega_b(6350)$ to be  the P-wave $\Omega_b$ states with the $J^P={\frac{3}{2}}^-$, ${\frac{1}{2}}^-$, ${\frac{5}{2}}^-$ and ${\frac{3}{2}}^-$, respectively.
\end{abstract}

 PACS number: 14.20.Mr

 Key words: $\Omega_b$,   QCD sum rules

\section{Introduction}
Recently, the LHCb collaboration reported four narrow peaks in the $\Xi_b^0K^-$ mass spectrum, the measured masses are
\begin{flalign}
M(\Omega_b(6316)) &= 6315.64\pm0.31\pm0.07\pm0.50 \,{\rm MeV}\, , \nonumber \\
M(\Omega_b(6330)) &= 6330.30\pm0.28\pm0.07\pm0.50 \,{\rm MeV}\, , \nonumber \\
M(\Omega_b(6340)) &= 6339.71\pm0.26\pm0.05\pm0.50 \,{\rm MeV}\, , \nonumber \\
M(\Omega_b(6350)) &= 6349.88\pm0.35\pm0.05\pm0.50 \,{\rm MeV}\, ,
\end{flalign}
 where the uncertainties are statistical, systematic and the last is due to the knowledge of the $\Xi_b^0$ mass \cite{Omegab-LHCb}.
The significances of the $\Omega_b(6316)$ and $\Omega_b(6330)$
peaks are  2.1$\sigma$ and 2.6$\sigma$ respectively, while the significances of the  $\Omega_b(6340)$ and $\Omega_b(6350)$ peaks exceed~5$\sigma$.

In the constituent quark models, the $\Omega_b$ states have three valence quarks $s$, $s$ and $b$. Without introducing an additional P-wave,
we obtain the ground state $\Omega_b$ states, $\Omega_b({\frac{1}{2}}^+)$ and $\Omega_b({\frac{3}{2}}^+)$, which have the spin-parity $J^P={\frac{1}{2}}^+$ and ${\frac{3}{2}}^+$, respectively.
Up to now, only the $\Omega_b({\frac{1}{2}}^+)$ is observed, the $\Omega_b({\frac{3}{2}}^+)$ has not been established yet \cite{PDG}. If there exists a relative P-wave between the two $s$-quarks or between the $ss$-diquark and $b$-quark, we obtain  five negative-parity $\Omega_b$ states. If exciting a P-wave costs about $300-350\,\rm{MeV}$, just like in the case of the $\Omega_c$ states, the P-wave $\Omega_b$ states should  have the masses about $6350-6400\,\rm{MeV}$. Direct calculations based  on the quark models and diquark-quark models indicate that the P-wave $\Omega_b$ baryon states have the masses about $6.30-6.50\,\rm{GeV}$ \cite{Oemgab-quark-model,WangNegativeP}.

In 2017, the LHCb collaboration studied the  $\Xi_c^+ K^-$  mass spectrum, and observed five new narrow excited $\Omega_c^0$ states,
$\Omega_c(3000)$, $\Omega_c(3050)$, $\Omega_c(3066)$, $\Omega_c(3090)$, $\Omega_c(3119)$ \cite{LHCb-Omega}.
If they are P-wave $\Omega_c$ states \cite{Omegac-LHCb,WZG-Omegac-negative}, the mass-gaps between the S-wave and P-wave $\Omega_c$ states are about $300-350\,\rm{MeV}$.
Other assignments, such as the 2S $\Omega_c$  states  with the spin-parity $J^P={\frac{1}{2}}^+$ and ${\frac{3}{2}}^+$ cannot be  excluded at the present time \cite{Azizi-Omega-2S}.

After the discovery of the $\Omega_b(6316)$,  $\Omega_b(6330)$,  $\Omega_b(6340)$ and $\Omega_b(6350)$, Chen et al studied the masses and decay widths of the P-wave $\Omega_b$ states via the QCD sum rules combined with the heavy quark effective theory \cite{ChenHX-Omegab}, while Liang and Lu studied the strong decays of
 those $\Omega_b$ states with the ${}^3P_0$ model and assigned  them as the $\lambda$-model P-wave $\Omega_b$ states \cite{Liang-Lu-Omegab}.

In previous work \cite{WZG-Omegac-negative}, we tentatively assigned  the $\Omega_c(3000)$, $\Omega_c(3050)$, $\Omega_c(3066)$, $\Omega_c(3090)$ and $\Omega_c(3119)$  to be the  P-wave $\Omega_c$ states with $J^P={\frac{1}{2}}^-$, ${\frac{1}{2}}^-$, ${\frac{3}{2}}^-$, ${\frac{3}{2}}^-$ and ${\frac{5}{2}}^-$, respectively, introduced a  relative P-wave explicitly in constructing the  current operators, and studied them with the full  QCD sum rules. In this article, we extend our previous work to study the $\Omega_b(6316)$,  $\Omega_b(6330)$,  $\Omega_b(6340)$ and $\Omega_b(6350)$ states as the P-wave $\Omega_b$  states  with the full QCD sum rules.

 The article is arranged as follows:  we derive the QCD sum rules for  the $\Omega_b^-$ states as P-wave baryons  in Sect.2;
 in Sect.3, we present the numerical results and discussions; and Sect.4 is reserved for our
conclusions.

\section{QCD sum rules for  the P-wave $\Omega_b^-$ states}

Now let us write down  the two-point correlation functions $\Pi(p)$, $\Pi_{\mu\nu}(p)$, $\Pi_{\mu\nu\alpha\beta}(p)$ firstly,
\begin{eqnarray}
\Pi(p)&=&i\int d^4x e^{ip \cdot x} \langle0|T\left\{J(x)\bar{J}(0)\right\}|0\rangle \, , \nonumber\\
\Pi_{\mu\nu}(p)&=&i\int d^4x e^{ip \cdot x} \langle0|T\left\{J_{\mu}(x)\bar{J}_{\nu}(0)\right\}|0\rangle \, , \nonumber\\
\Pi_{\mu\nu\alpha\beta}(p)&=&i\int d^4x e^{ip \cdot x} \langle0|T\left\{J_{\mu\nu}(x)\bar{J}_{\alpha\beta}(0)\right\}|0\rangle \, ,
\end{eqnarray}
where  $J_\mu(x)=J^1_\mu(x),\,J^2_\mu(x)$,
\begin{eqnarray}
J(x)&=&i\varepsilon^{ijk} \left[ \partial^\mu s^T_i(x) C\gamma^\nu s_j(x)+ s^T_i(x) C\gamma^\nu \partial^{\mu}s_j(x)\right]\sigma_{\mu\nu}\,b_k(x) \, ,\nonumber \\
J^1_{\mu}(x)&=&i\varepsilon^{ijk} \left[ \partial^\alpha s^T_i(x) C\gamma^\beta s_j(x)+s^T_i(x) C\gamma^\beta \partial^{\alpha}s_j(x)\right]\left(\widetilde{g}_{\mu\alpha}\gamma_\beta-\widetilde{g}_{\mu\beta}\gamma_\alpha \right)\gamma_5 b_k(x) \, ,\nonumber \\
J^2_{\mu}(x)&=&i\varepsilon^{ijk} \left[ \partial^\alpha s^T_i(x) C\gamma^\beta s_j(x)+ s^T_i(x) C\gamma^\beta \partial^{\alpha}s_j(x)\right]\nonumber\\
&&\left(g_{\mu\alpha}\gamma_\beta+g_{\mu\beta}\gamma_\alpha-\frac{1}{2}g_{\alpha\beta}\gamma_\mu \right)\gamma_5 b_k(x) \, ,\nonumber\\
J_{\mu\nu}(x)&=&i\varepsilon^{ijk} \left[ \partial_\mu s^T_i(x) C\gamma_\nu s_j(x)+\partial_\nu s^T_i(x) C\gamma_\mu s_j(x)+ s^T_i(x) C\gamma_\nu \partial_{\mu}s_j(x)\right.\nonumber\\
&&\left.+ s^T_i(x) C\gamma_\mu \partial_{\nu}s_j(x)\right] b_k(x) \, ,
\end{eqnarray}
$\widetilde{g}_{\mu\nu}=g_{\mu\nu}-\frac{1}{4}\gamma_\mu\gamma_\nu$,
the $i$, $j$, $k$ are color indices, the $C$ is the charge conjugation matrix.
We choose the current operators $J(x)$, $J_\mu(x)$ and $J_{\mu\nu}(x)$ to study the P-wave $\Omega_b$ states with the spin $J=\frac{1}{2}$, $\frac{3}{2}$ and $\frac{5}{2}$, respectively. For detailed discussions on how to construct those current operators, one can consult Ref.\cite{WZG-Omegac-negative}.

 The current operators $J(0)$, $J_\mu(0)$ and $J_{\mu\nu}(0)$ couple potentially to the spin-parity  $J^P={\frac{1}{2}}^\mp$,  ${\frac{3}{2}}^\mp$ and ${\frac{5}{2}}^\mp$  $\Omega_b$ baryon
 states $\Omega_{\frac{1}{2}}^{\mp}$,  $\Omega_{\frac{3}{2}}^{\mp}$ and $\Omega_{\frac{5}{2}}^{\mp}$, respectively,
\begin{eqnarray}\label{Pole-residue-N}
\langle 0| J (0)|\Omega_{\frac{1}{2}}^{-}(p)\rangle &=&\lambda^{-}_{\frac{1}{2}} U^{-}(p,s) \, , \nonumber \\
\langle 0| J_{\mu} (0)|\Omega_{\frac{3}{2}}^{-}(p)\rangle &=&\lambda^{-}_{\frac{3}{2}} U^{-}_\mu(p,s) \, ,  \nonumber\\
\langle 0| J_{\mu\nu} (0)|\Omega_{\frac{5}{2}}^{-}(p)\rangle &=&\lambda^{-}_{\frac{5}{2}} U^{-}_{\mu\nu}(p,s) \, ,
\end{eqnarray}
 \begin{eqnarray}\label{Pole-residue-P}
 \langle 0| J (0)|\Omega_{\frac{1}{2}}^{+}(p)\rangle &=&\lambda^{+}_{\frac{1}{2}}  i\gamma_5 U^{+}(p,s) \, ,\nonumber  \\
\langle 0| J_{\mu} (0)|\Omega_{\frac{3}{2}}^{+}(p)\rangle &=&\lambda^{+}_{\frac{3}{2}}i\gamma_5 U^{+}_{\mu}(p,s) \, , \nonumber\\
\langle 0| J_{\mu\nu} (0)|\Omega_{\frac{5}{2}}^{+}(p)\rangle &=&\lambda^{+}_{\frac{5}{2}}i\gamma_5 U^{+}_{\mu\nu}(p,s) \, ,
\end{eqnarray}
as multiplying $i\gamma_5$ to the current operators can change their parity \cite{Oka96,WangPc,WangHbaryon}, where the $U^\pm(p,s)$, $U^{\pm}_\mu(p,s)$ and $U^{\pm}_{\mu\nu}(p,s)$ are Dirac and Rarita-Schwinger spinors, respectively,   the $\lambda^{\pm}_{j}$ with $j=\frac{1}{2}$, $\frac{3}{2}$, $\frac{5}{2}$ are the pole residues. For the properties of those  spinors, one can consult Refs.\cite{WZG-Omegac-negative,WangPc}.

At the hadron side,  we insert a complete set  of intermediate $\Omega_b$  states with same quantum numbers as the current operators $J(x)$,
$i\gamma_5 J(x)$, $J_\mu(x)$, $i\gamma_5 J_\mu(x)$, $J_{\mu\nu}(x)$ and
$i\gamma_5 J_{\mu\nu}(x)$ into the correlation functions, and
 take into account the  possible  current-baryon couplings defined in Eqs.\eqref{Pole-residue-N}-\eqref{Pole-residue-P} to obtain the hadron representation,
\begin{eqnarray}
 \Pi(p) & = & {\lambda^{-}_{\frac{1}{2}}}^2  {\!\not\!{p}+ M_{-} \over M_{-}^{2}-p^{2}  }
 +  {\lambda^{+}_{\frac{1}{2}}}^2  {\!\not\!{p}- M_{+} \over M_{+}^{2}-p^{2}  } +\cdots  \, , \nonumber\\
 &=&\Pi_{\frac{1}{2}}(p^2)\, ,
 \end{eqnarray}
 \begin{eqnarray}
  \Pi_{\mu\nu}(p) & = & \left[{\lambda^{-}_{\frac{3}{2}}}^2  {\!\not\!{p}+ M_{-} \over M_{-}^{2}-p^{2}  }
+  {\lambda^{+}_{\frac{3}{2}}}^2  {\!\not\!{p}- M_{+} \over M_{+}^{2}-p^{2}  } \right] \left(- g_{\mu\nu}+\cdots\right) +\cdots  \, ,\nonumber\\
&=&\Pi_{\frac{3}{2}}(p^2)\,\left(- g_{\mu\nu}\right)+\cdots\, ,
\end{eqnarray}
\begin{eqnarray}
\Pi_{\mu\nu\alpha\beta}(p) & = & \left[{\lambda^{-}_{\frac{5}{2}}}^2  {\!\not\!{p}+ M_{-} \over M_{-}^{2}-p^{2}  }+{\lambda^{+}_{\frac{5}{2}}}^2  {\!\not\!{p}- M_{+} \over M_{+}^{2}-p^{2}  }\right] \left(\frac{ \widetilde{g}_{\mu\alpha}\widetilde{g}_{\nu\beta}}{2}+ \cdots\right) +\cdots \, ,\nonumber\\
&=&\Pi_{\frac{5}{2}}(p^2)\,\frac{ g_{\mu\alpha}g_{\nu\beta}+g_{\mu\beta}g_{\nu\alpha}}{2}+\cdots \, ,
\end{eqnarray}
where $\widetilde{g}_{\mu\nu}=g_{\mu\nu}-\frac{p_{\mu}p_{\nu}}{p^2}$. We choose the tensor structures $g_{\mu\nu}$ and $g_{\mu\alpha}g_{\nu\beta}+g_{\mu\beta}g_{\nu\alpha}$ to study the spin $j=\frac{3}{2}$ and $\frac{5}{2}$ $\Omega_b$ states, respectively  \cite{WZG-Omegac-negative,WangPc}.

Now it is straightforward to get  the hadronic spectral densities  through dispersion relation,
\begin{eqnarray}
\frac{{\rm Im}\Pi_{j}(s)}{\pi}&=&\!\not\!{p} \left[{\lambda^{-}_{j}}^2 \delta\left(s-M_{-}^2\right)+{\lambda^{+}_{j}}^2 \delta\left(s-M_{+}^2\right)\right] +\left[M_{-}{\lambda^{-}_{j}}^2 \delta\left(s-M_{-}^2\right)-M_{+}{\lambda^{+}_{j}}^2 \delta\left(s-M_{+}^2\right)\right]\, , \nonumber\\
&=&\!\not\!{p}\, \rho^1_{j,H}(s)+\rho^0_{j,H}(s) \, ,
\end{eqnarray}
where $j=\frac{1}{2}$, $\frac{3}{2}$, $\frac{5}{2}$, we add the subscript $H$ to represent the hadron side.

At the QCD side, we  carry out the operator product expansion up to the vacuum condensates of dimension 10 and take into account  the vacuum condensates $\langle\bar{s}s\rangle$, $\langle \frac{\alpha_sGG}{\pi}\rangle$, $\langle\bar{s}g_s\sigma Gs\rangle$, $\langle\bar{s}s\rangle\langle\bar{s}g_s\sigma Gs\rangle$, $\langle\bar{s}g_s\sigma Gs\rangle^2$, which are vacuum expectations of the quark-gluon operators of the order $\mathcal{O}(\alpha_s^k)$ with $k\leq1$, the vacuum condensate $\langle\bar{s}s\rangle^2$ has no contribution. Again, we get  the QCD spectral densities  through  dispersion relation,
\begin{eqnarray}
\frac{{\rm Im}\Pi_{j}(s)}{\pi}&=&\!\not\!{p}\, \rho^1_{j,QCD}(s)+\rho^0_{j,QCD}(s) \, ,
\end{eqnarray}
where $j=\frac{1}{2}$, $\frac{3}{2}$, $\frac{5}{2}$, the interested readers can acquire the explicit expressions of the QCD spectral densities $\rho^1_{j,QCD}(s)$ and $\rho^0_{j,QCD}(s)$ via contacting me with  E-mail.

Now let us  implement  the quark-hadron duality below the continuum thresholds  $s_0$ and introduce the weight function $\exp\left(-\frac{s}{T^2}\right)$ to get   the  QCD sum rules:
\begin{eqnarray}\label{QCDSR}
2M_{-}{\lambda^{-}_{j}}^2\exp\left( -\frac{M_{-}^2}{T^2}\right)
&=& \int_{m_c^2}^{s_0}ds \left[\sqrt{s}\rho^1_{j,H}(s)+\rho^0_{j,H}(s)\right]\exp\left( -\frac{s}{T^2}\right)\nonumber\\
&=&\int_{m_c^2}^{s_0}ds \left[\sqrt{s}\rho^1_{j,QCD}(s)+\rho^0_{j,QCD}(s)\right]\exp\left( -\frac{s}{T^2}\right)\, ,
\end{eqnarray}
where the $T^2$ is the Borel parameter.

We derive   Eq.\eqref{QCDSR} in regard  to  $\frac{1}{T^2}$, then eliminate the
 pole residues $\lambda^{-}_j$ and get  the QCD sum rules for
 the masses of the $\Omega_b$  states with negative-parity,
 \begin{eqnarray}
 M^2_{-} &=& \frac{-\frac{d}{d(1/T^2)}\int_{m_c^2}^{s_0}ds \left[\sqrt{s}\rho^1_{j,QCD}(s)+\rho^0_{j,QCD}(s)\right]\exp\left( -\frac{s}{T^2}\right)}{\int_{m_c^2}^{s_0}ds \left[\sqrt{s}\rho^1_{j,QCD}(s)+\rho^0_{j,QCD}(s)\right]\exp\left( -\frac{s}{T^2}\right)}\, .
\end{eqnarray}

\section{Numerical results and discussions}
We choose   the standard values of the vacuum condensates
$\langle\bar{q}q \rangle=-(0.24\pm 0.01\, \rm{GeV})^3$,  $\langle\bar{s}s \rangle=(0.8\pm0.1)\langle\bar{q}q \rangle$,
 $\langle\bar{s}g_s\sigma G s \rangle=m_0^2\langle \bar{s}s \rangle$,
$m_0^2=(0.8 \pm 0.1)\,\rm{GeV}^2$, $\langle \frac{\alpha_s
GG}{\pi}\rangle=(0.33\,\rm{GeV})^4 $    at the energy scale  $\mu=1\, \rm{GeV}$
\cite{SVZ79,PRT85,ColangeloReview}, the $\overline{MS}$ masses $m_{b}(m_b)=(4.18\pm0.03)\,\rm{GeV}$ and $m_s(\mu=2\,\rm{GeV})=(0.095\pm0.005)\,\rm{GeV}$
 from the Particle Data Group \cite{PDG}. We extract the masses of the P-wave $\Omega_b$ states at the best energy scales $\mu$ of the QCD spectral densities, the input parameters  evolve with the energy scale $\mu$ according to the re-normalization  group equation,
\begin{eqnarray}
\langle\bar{s}s \rangle(\mu)&=&\langle\bar{s}s \rangle({\rm 1GeV})\left[\frac{\alpha_{s}({\rm 1GeV})}{\alpha_{s}(\mu)}\right]^{\frac{12}{23}}\, , \nonumber\\
 \langle\bar{s}g_s \sigma G s \rangle(\mu)&=&\langle\bar{s}g_s \sigma Gs \rangle({\rm 1GeV})\left[\frac{\alpha_{s}({\rm 1GeV})}{\alpha_{s}(\mu)}\right]^{\frac{2}{23}}\, , \nonumber\\
m_b(\mu)&=&m_b(m_b)\left[\frac{\alpha_{s}(\mu)}{\alpha_{s}(m_b)}\right]^{\frac{12}{23}} \, ,\nonumber\\
m_s(\mu)&=&m_s({\rm 2GeV} )\left[\frac{\alpha_{s}(\mu)}{\alpha_{s}({\rm 2GeV})}\right]^{\frac{12}{23}} \, ,\nonumber\\
\alpha_s(\mu)&=&\frac{1}{b_0t}\left[1-\frac{b_1}{b_0^2}\frac{\log t}{t} +\frac{b_1^2(\log^2{t}-\log{t}-1)+b_0b_2}{b_0^4t^2}\right]\, ,
\end{eqnarray}
   where $t=\log \frac{\mu^2}{\Lambda_{QCD}^2}$, $b_0=\frac{33-2n_f}{12\pi}$, $b_1=\frac{153-19n_f}{24\pi^2}$, $b_2=\frac{2857-\frac{5033}{9}n_f+\frac{325}{27}n_f^2}{128\pi^3}$,  $\Lambda_{QCD}=210\,\rm{MeV}$, $292\,\rm{MeV}$  and  $332\,\rm{MeV}$ for the flavors  $n_f=5$, $4$ and $3$, respectively  \cite{PDG,Narison-mix},  we take the flavor $n_f=5$.

In Refs.\cite{WangTetraquark,WangMolecule}, we study the  energy scale dependence  of the QCD sum rules  for the hidden-charm (bottom) tetraquark states and molecular  states  for the first time,  and suggest an  energy scale  formula $\mu=\sqrt{M^2_{X/Y/Z}-(2{\mathbb{M}}_Q)^2}$ to choose  the best  energy scales of the QCD spectral densities, where the $X$, $Y$, $Z$ are  the four-quark (exotic) states, and the ${\mathbb{M}}_Q$ are the effective heavy quark masses or constituent quark masses.
  If we resort to  the diquark-quark model to construct the current operators to interpolate the heavy baryon states $B$, and  there exists an analogous energy scale formula
$ \mu =\sqrt{M_{B}^2-{\mathbb{M}}_Q^2}$
   to choose the  best energy scales of the QCD spectral densities \cite{WZG-Omegac-negative}.  We choose  the updated value ${\mathbb{M}}_b=5.17\,\rm{GeV}$ fitted in the QCD sum rules for the diquark-antidiquark type hidden-bottom tetraquark states \cite{WangEPJC-Hiddenbottom}, then $\mu=\sqrt{6.35^2-5.17^2}\,\rm{GeV}\approx3.7\,\rm{GeV}$. In this article,  we set the energy scales  to be $\mu=3.7\,\rm{GeV}$.

Now let us  search for the best Borel parameters and continuum threshold parameters to warrant convergence of the operator product expansion at the QCD side and  pole dominance at the hadron side via trial  and error. Finally, we get the  Borel windows $T^2$, continuum threshold parameters $s_0$,  pole contributions (or the contributions below the continuum thresholds $s_0$)  and  perturbative contributions (or the contributions of  the perturbative terms), see Table \ref{Borel-pole-per}.
From the Table, we observe that the pole contributions are about $(40-60)\%$, the pole dominance criterion is satisfied, on the other hand, the main contributions come from the perturbative terms, the operator product expansion converges very good.

\begin{table}
\begin{center}
\begin{tabular}{|c|c|c|c|c|c|}\hline\hline
currents      & $T^2 (\rm{GeV}^2)$   & $\sqrt{s_0} (\rm{GeV})$    & Pole          & Pert \\  \hline
$J^1_\mu$     & $5.0-5.6$            & $7.0\pm0.1$                & $(42-61)\%$   & $(90-92)\%$ \\ \hline
$J$           & $4.9-5.5$            & $7.0\pm0.1$                & $(42-62)\%$   & $(90-92)\%$ \\ \hline
$J_{\mu\nu}$  & $5.2-5.8$            & $7.0\pm0.1$                & $(42-61)\%$   & $(96-97)\%$ \\ \hline
$J^2_\mu$     & $5.0-5.6$            & $7.0\pm0.1$                & $(41-61)\%$   & $(98-99)\%$ \\ \hline\hline
\end{tabular}
\end{center}
\caption{ The Borel windows $T^2$, continuum threshold parameters $s_0$,
 pole contributions (Pole)   and  perturbative contributions (Pert).}\label{Borel-pole-per}
\end{table}

\begin{table}
\begin{center}
\begin{tabular}{|c|c|c|c|c|c|c|}\hline\hline
 currents       & $J_{j_l}^P$               & $M_\Omega(\rm{GeV})$   &$\lambda_\Omega (10^{-1}\rm{GeV}^4)$   & assignments \\ \hline
 $J^1_\mu$      & ${\frac{3}{2}}_{1}^-$     & $6.31\pm0.11$          &$1.25\pm0.21$                          & $\Omega_b(6316)$\\ \hline
  $J$           & ${\frac{1}{2}}_{0}^-$     & $6.32\pm0.11$          &$2.80\pm0.47$                          & $\Omega_b(6330)$ \\ \hline
  $J_{\mu\nu}$  & ${\frac{5}{2}}_{2}^-$     & $6.35\pm0.10$          &$2.46\pm0.37$                          & $\Omega_b(6340)$ \\ \hline
 $J^2_\mu$      & ${\frac{3}{2}}_{2}^-$     & $6.37\pm0.09$          &$4.28\pm0.66$                          & $\Omega_b(6350)$ \\ \hline  \hline
\end{tabular}
\end{center}
\caption{  The  masses $M_\Omega$,  pole residues $\lambda_\Omega$ and possible assignments of the $\Omega_b$ states, where the $j_l$ is the total angular momentum of the light degree of freedom.}\label{mass-assign}
\end{table}

\begin{figure}
 \centering
 \includegraphics[totalheight=5cm,width=7cm]{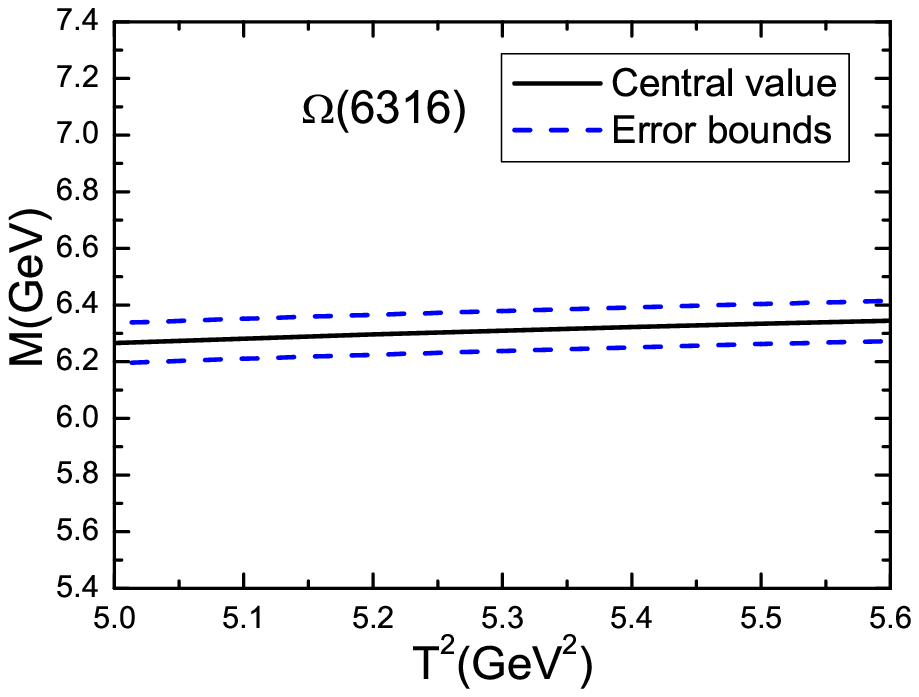}
 \includegraphics[totalheight=5cm,width=7cm]{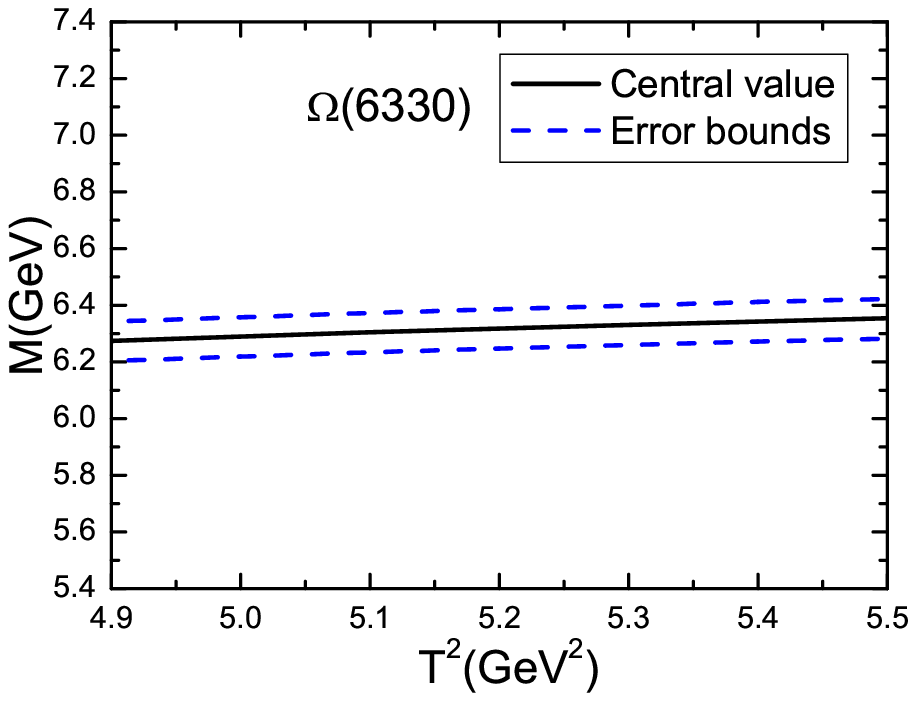}
 \includegraphics[totalheight=5cm,width=7cm]{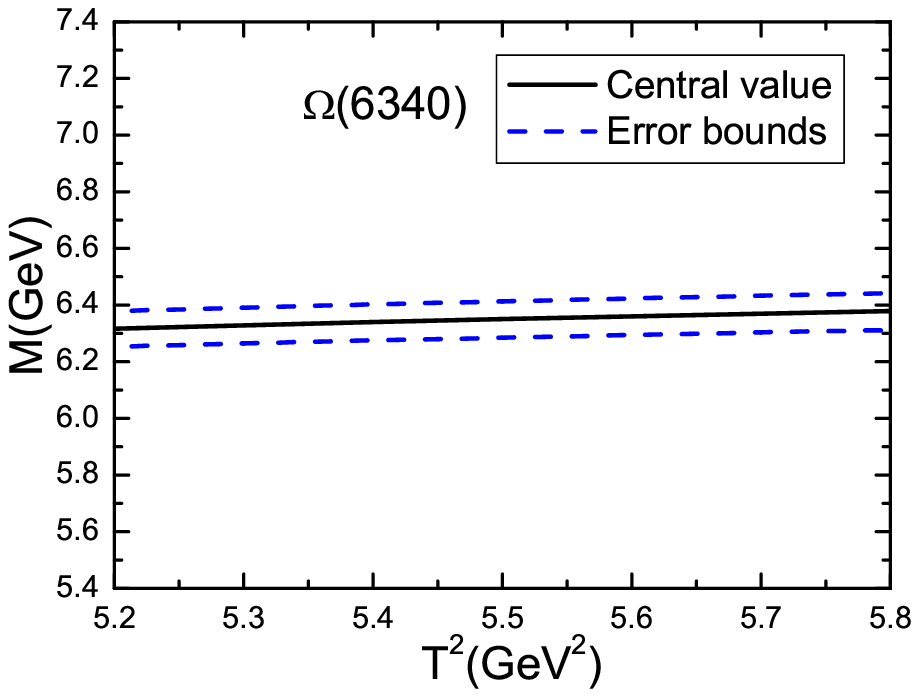}
 \includegraphics[totalheight=5cm,width=7cm]{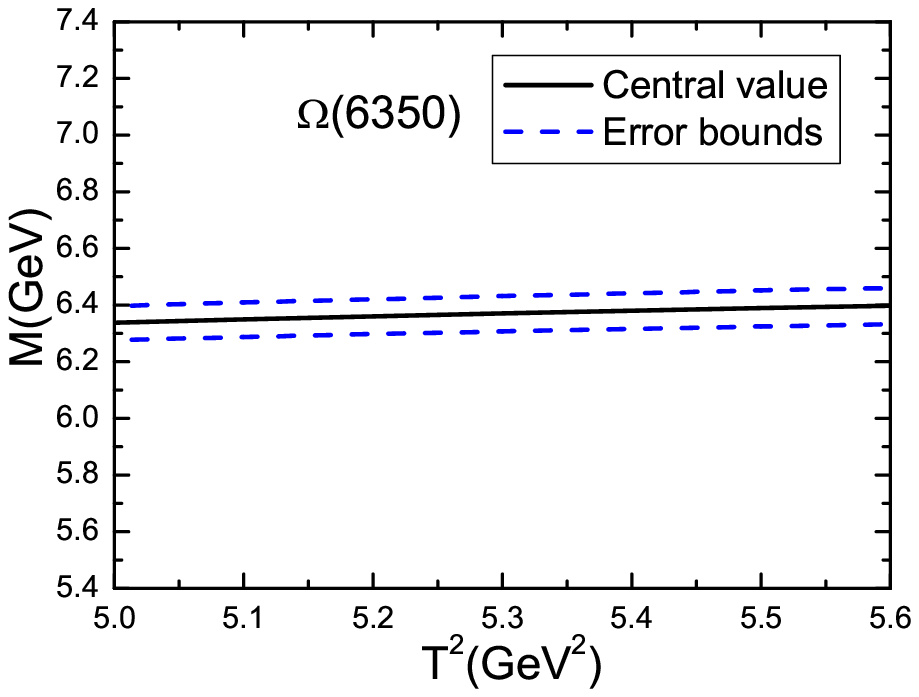}
       \caption{ The masses  of the P-wave $\Omega_b$ states  with variations of the Borel parameters $T^2$.  }\label{mass-Borel}
\end{figure}

Now we take into account all uncertainties  of the input  parameters, and get  the values of the masses and pole residues of the P-wave $\Omega_b$ states, which are shown in Fig.\ref{mass-Borel} and Table \ref{mass-assign}. In Fig.\ref{mass-Borel},  we plot the masses of the P-wave $\Omega_b$ states with variations
of the Borel parameters $T^2$. From the four diagrams in the figure, we can see that there appear very flat platforms indeed,  the uncertainties come from the Borel parameters $T^2$ are very small,
it is reliable to extract the  masses of the P-wave $\Omega_b$ states.

From Table \ref{mass-assign}, we can see that the predicted masses of the  P-wave $\Omega_b$ states are all consistent with the experimental values of the masses of the
$\Omega_b(6316)$, $\Omega_b(6330)$, $\Omega_b(6340)$ and $\Omega_b(6350)$ from the LHCb collaboration within uncertainties \cite{Omegab-LHCb}.
The central values of the masses $6.31\pm0.11\,\rm{GeV}$, $6.32\pm0.11\,\rm{GeV}$, $6.35\pm0.10\,\rm{GeV}$ and
$6.37\pm0.09\,\rm{GeV}$ shown in Table \ref{mass-assign} come from the QCD sum rules with the same continuum threshold parameters $\sqrt{s_0}=7.0\,\rm{GeV}$, pole contributions $51\%$, and energy scales
of the QCD spectral densities $\mu=3.7\,\rm{GeV}$. We can get the conclusion tentatively that the predicted masses of the P-wave $\Omega_b$ states have the hierarchy
   $M_{J^1_\mu}<M_{J}<M_{ J_{\mu\nu}}<M_{J^2_\mu}$, where we use the currents to represent  the corresponding $\Omega_b$ states.     The present calculations favor assigning the $\Omega_b(6316)$, $\Omega_b(6330)$, $\Omega_b(6340)$ and $\Omega_b(6350)$ as the P-wave $\Omega_b$ states with the spin-parity $J^P={\frac{3}{2}}^-$, ${\frac{1}{2}}^-$, ${\frac{5}{2}}^-$ and ${\frac{3}{2}}^-$, respectively, see Table \ref{mass-assign}.

The LHCb collaboration observed the four narrow structures $\Omega_b(6316)$, $\Omega_b(6330)$, $\Omega_b(6340)$ and $\Omega_b(6350)$ in the $\Xi_b^0K^-$ mass spectrum. In this article, we study  the P-wave $\Omega_b$ states, which have an explicit P-wave between the two $s$-quarks. The decays $\Omega_b(6316/6330/6340/6350)\to\Xi_b^0K^-$ take place by
creating a $u\bar{u}$ pair with $J^{P}=0^+$  from the QCD vacuum,  the relative P-wave between the two $s$-quarks frustrates the formation of the S-wave scalar $us$-diquark correlation so as to form the $\Xi_b^0$ baryon, which can account for the narrow  widths of the $\Omega_b$ states.

In Ref.\cite{WangNegativeP}, we choose the currents without introducing relative P-waves to study the negative parity heavy and doubly-heavy   baryon states in an  systematic way, and obtain the predictions for the masses $M=2.98\pm0.16 \,\rm{GeV}$ and $6.27\pm 0.14\,\rm{GeV}$   for the $J^P={\frac{1}{2}}^-$ heavy baryon states $\Omega_c^0$ and $\Omega_b^{-}$, respectively, where the diquark constituent or operator
$\varepsilon^{ijk}s^T_jC\gamma_\mu s_k$ is chosen to construct the interpolating currents $\tilde{J}(x)$ with the spin-parity $J^P={\frac{1}{2}}^+$,
\begin{eqnarray}
\tilde{J}(x)&=&\varepsilon^{ijk}s^T_i(x)C\gamma_\mu s_j(x)\gamma^{\mu}\gamma_5Q_k(x)\, ,
\end{eqnarray}
  $Q=c$ and $b$. As multiplying $i \gamma_{5}$ to the baryon currents $\tilde{J}(x)$ changes their parity, we can also choose the currents  without introducing relative P-waves to study the P-wave baryon states.
The currents $\tilde{J}(x)$ also couple potentially to the $\Omega_c^0$ or $\Omega_b^-$ state with the spin-parity $J^P={\frac{1}{2}}^-$ \cite{WangNegativeP}, the  mass of the $\Omega_c(3000)$ from the LHCb collaboration is in very good  agreement with the prediction $M=2.98\pm0.16 \,\rm{GeV}$ from the QCD sum rules \cite{WangNegativeP}. In Ref.\cite{WZG-Omegac-negative}, we assign the $\Omega_c(3050)$, $\Omega_c(3066)$, $\Omega_c(3090)$ and $\Omega_c(3119)$  to be the  P-wave baryon states with the spin-parity $J^P={\frac{1}{2}}^-$, ${\frac{3}{2}}^-$, ${\frac{3}{2}}^-$ and ${\frac{5}{2}}^-$, respectively, where the two $s$ quarks are in relative P-wave; and assign  the  $\Omega_c(3000)$  to be the P-wave baryon state with the spin-parity $J^{P}={\frac{1}{2}}^-$, where the two $s$ quarks are in relative S-wave.
At the present time,  there is no experimental candidate for the
corresponding $\Omega_b^-$ state with the spin-parity $J^P={\frac{1}{2}}^-$, where the two $s$ quarks are in relative S-wave.
In Ref.\cite{WZG-Omegac-negative}, we also construct the current $\hat{J}(x)$ with the spin-parity $J^P={\frac{1}{2}}^-$,
\begin{eqnarray}
\hat{J}(x)&=&i\varepsilon^{ijk} \left[ \partial^\mu s^T_i(x) C\gamma^\nu s_j(x)+ s^T_i(x) C\gamma^\nu \partial^{\mu}s_j(x)\right]g_{\mu\nu}\,c_k(x)\, ,
\end{eqnarray}
to study the negative parity $\Omega_c^0$ states, but cannot obtain stable QCD sum rules. In this article, we abandon the corresponding current $\hat{J}(x)\mid_{c\to b}$.

\section{Conclusion}
In this article, we introduce an explicit P-wave between the two $s$-quarks to construct the current operators to study the P-wave $\Omega_b$ states with the full  QCD sum rules  by carrying out the operator product expansion up to the vacuum condensates of dimension $10$. In calculations, we resort to the energy scale  formula to choose the best energy scales  of the QCD spectral densities, and get very stable QCD sum rules in the Borel widows, where the operator product expansion converges very good and the contributions of the pole terms are satisfactory. The present calculations favor assigning the $\Omega_b(6316)$, $\Omega_b(6330)$, $\Omega_b(6340)$ and $\Omega_b(6350)$ to be  the P-wave $\Omega_b$ states with the $J^P={\frac{3}{2}}^-$, ${\frac{1}{2}}^-$, ${\frac{5}{2}}^-$ and ${\frac{3}{2}}^-$, respectively.

\section*{Acknowledgements}
This  work is supported by National Natural Science Foundation, Grant Number  11775079.

\end{document}